\documentclass[letterpaper, 10 pt, conference]{ieeeconf}
\usepackage{enumerate}
\usepackage{courier}
\usepackage{hyperref}
\usepackage{graphicx,subfigure}
\usepackage{epstopdf}
\usepackage{tikz}
\usepackage{amsmath}
\usepackage{multirow}
\usepackage{amsfonts}
\usepackage{float}
\usepackage{amssymb}
\usepackage{graphicx}

\usepackage{algorithm,float}
\usepackage{algpseudocode}
\usepackage{verbatim}
\newtheorem{theorem}{Theorem}
\newtheorem{lemma}{Lemma}

\newtheorem{corollary}{Corollary}
\newtheorem{definition}{Definition}
\newtheorem{remark}{Remark}
\newtheorem{example}{Example}

\makeatletter

\newcommand{\Rmnum}[1]{\expandafter\@slowromancap\romannumeral #1@}

\title{\LARGE \bf
Neuromimetic Dynamic Networks with Hebbian Learning
}
\author{Zexin Sun \& John Baillieul}
\makeatother
\begin{document}

\maketitle

	\let\thefootnote\relax\footnotetext{\noindent\underbar{\hspace{3.25in}}\\
		
		Zexin Sun is with the Division of Systems Engineering at Boston University, Boston, MA 02215, {(email: \tt zxsun@bu.edu}).
	
	John Baillieul is with the Departments of Mechanical Engineering, Electrical and Computer Engineering, and the Division of Systems Engineering at Boston University, Boston, MA 02215, {(email: \tt johnb@bu.edu}).}
\begin{abstract}
Leveraging recent advances in neuroscience and control theory, this paper presents a neuromimetic network model with dynamic symmetric connections governed by Hebbian learning rules. Formal analysis grounded in graph theory and classical control establishes that this biologically plausible model exhibits boundedness, stability, and structural controllability given a generalized sym-cactus structure with multiple control nodes. We prove the necessity of this topology when there are distributed control inputs. Simulations using a 14-node generalized sym-cactus network with two input types validate the model’s effectiveness in capturing key neural dynamics.
\end{abstract}

\section{Introduction}

The astounding breadth and complexity of animal behavior, from basic motor functions to advanced cognition, emerges from the interconnected neural networks distributed throughout their bodies. Cross-disciplinary research continues eo elucidate the way that neural structure and dynamics give rise to perception, thought, and action. On the simpler end of the spectrum, hermaphrodite nematodes ({\em C. elegans}) contain a mere 302 neurons with around 460 synaptic connections yet still carry out surprisingly sophisticated behaviors governed by intricate neurophysiological processes that remain poorly understood and have yet to be faithfully replicated in even the most advanced artificial systems. By contrast, the human nervous system contains upwards of 86 billion neurons, with approximately 0.15 quadrillion synapses, manifesting the apex of biological neural complexity \cite{cook2019whole}. However, while nematodes have orders of magnitude fewer neurons, their underlying neurophysiology still produces multifaceted capabilities far surpassing any current computational models. Since the pioneering work of McCulloch and Pitts formalized neurons as logical computing elements \cite{mcculloch1943logical}, substantial research has gradually assembled pieces of the neurobiology puzzle. Significant discoveries include the characterization of single neuron dynamics, beginning with the work of Hodgkin and Huxley on action potential generation \cite{hodgkin1952currents}. Later advances illuminated emergent computations in interconnected neural networks, including Hopfield's foundational recurrent neural network model \cite{hopfield1982neural}. Careful study of biological learning mechanisms has also informed more robust artificial neural network training techniques, providing alternatives to fragile backpropagation-based approaches. Recent work has developed unsupervised, interpretable models matching performance on tasks like MNIST digit classification \cite{Krotov2019}, beginning to bridge neuroscience and machine learning. However, vast gaps remain in our understanding of how brains produce minds and how neural computation gives rise to cognition and behavior. Ongoing interdisciplinary efforts are gradually advancing our understanding of both natural and artificial neural systems.

A longstanding focus on individual neural components remains crucial, but network-level modeling promises new insights into both neurological function and dysfunction. Ongoing efforts to bridge network, cellular, and molecular neuroscience with advanced computing may enable truly brain-like artificial intelligence while also elucidating the astonishing capabilities of natural neural systems. Recent research reveals the importance of brain-wide structural connectivity in cognition and dysfunction. In particular, network science has illuminated how largescale brain topology relates to neurological disorders like Alzheimer's disease  \cite{Gaiteri2016} and amnesia \cite{Ryan2022}. For instance, experiments demonstrate how engram cell accessibility changes in amnesia arise from alterations in cell-specific synaptic weights  \cite{Gaiteri2016}. To probe such networked dynamics, recent modeling work leverages tools from control theory and graph theory to capture how cognitive brain networks transition between states, \cite{Gu2015}. Our own prior work on neuromimetic control was aimed at understanding the dynamic characteristics of control systems actuated by the aggregate effects of large numbers of binary inputs — simulating neural states of spiking or resting \cite{Baillieul2019,Sun2022,Sun2023}. Using algorithms similar to Ola’s rule, it was shown how our neuromimetic models could emulate classical control design produced by small numbers of analogue inputs. Other work on Hebbian type learning \cite{Veronica2022} has shown how features of predictive cognition emerge as training progresses.
Ongoing efforts continue advancing explainable network-level models of learning, memory, and higher-level cognition \cite{Sherrill2015,Ryan2022,Roy2016}. In these efforts, both detailed neuronal networks and abstracted regional connectivity graphs prove useful, providing complementary perspectives. This paper contributes a coupled pre/post-synaptic model undergoing Hebbian plasticity, capturing salient aspects of synaptic learning for excitatory and inhibitory connections. Analysis via graph and control theory derives conditions guaranteeing consistency with known neuronal features, such as controllability \cite{Gu2015} and synaptic plasticity \cite{Abb2000}. The results will be validated by simulations that show the model's biological fidelity.

The rest of this paper is organized as follows. Section II presents preliminary notations and motivations for the proposed model. Section III provides formal analysis across three model features: boundedness, stability, and structural controllability. Section IV demonstrates via simulations that the model exhibits biologically plausible dynamics. Finally, Section V summarizes the key contributions and results.

\section{Modelling of Neuromimetic Dynamic Networks}

We propose a dynamic neural network model capturing both individual neuron dynamics and network-level features of biological neuronal connectomes. Key aspects include synaptic weight evolution per the Hebbian learning rule and interconnections between coupled brain regions. Most work such as the Hopfield network \cite{hopfield1982neural} and the linearized Wilson–Cowan system \cite{Galan2008} assume brain connections can be viewed as symmetric directed graphs. In this context, we construct a graph $\mathcal{G} = (\mathcal{V}, \mathcal{E})$ where $\mathcal{V}$ denotes the vertex set of brain regions and $\mathcal{E} \subseteq \mathcal{V} \times \mathcal{V}$ denotes brain regions and white matter connections, respectively. $N = |\mathcal{V}|$ gives the total number of nodes. Edges are partitioned into positive and negative weight sets $\mathcal{E}^{+}$ and $\mathcal{E}^{-}$, with $\mathcal{E} = \mathcal{E}^{+} \cup \mathcal{E}^{-}$. The weight matrix $A=[a_{ij}]$ has entries $a_{ij}$ representing the connection strength from node $j$ to node $i$. We define $d_{max}$ as the maximum out-degree over the directed graph $\mathcal{G}$.


The whole model is constructed as:
\begin{subequations}
	\label{eq:model_1}
	\begin{align}
	\dot x_{i}(t)&=-c_nx_{i}(t)+\sum_{(j,i)\in\mathcal{E}}a_{ij}(t)x_{j}(t)+ b_iu_{i},\label{1a}\\
	\begin{split}
	a_{ij}(t+\tau)&=\left[c_a^-a_{ij}(t)+\phi(x_i(t)x_j(t))\right]_{\underline a^-}^{\overline a^-}, (j,i)\in\mathcal{E}^-,\label{1b}\\
	a_{ij}(t+\tau)&=\left[c_a^+a_{ij}(t)+\phi(x_i(t)x_j(t))\right]_{\underline a^+}^{\overline a^+}, (j,i)\in\mathcal{E}^+,
	\end{split}
	\end{align}
\end{subequations}
where $i,j\in \mathcal{V}$ and $\tau$ is a time constant for weight updates. Symmetry requires $a_{ij} = a_{ji}$ and $(i,j) \in \mathcal{E} \Rightarrow (j,i) \in \mathcal{E}$. Equation (\ref{1a})is a continuous-time Hopfield network, with $b_i=0$ indicating no external input to node $i$. Equation (\ref{1b}) models region interconnectivity that changes more slowly than neural state, per synaptic plasticity strengthening/weakening rules \cite{Citri2008}. Over short time $\tau$, synapse strength is constant. For any $(j,i)$ pair, it can be written as
\begin{equation}
	\label{amp2}
	a_{ij}(t)=\left\{\begin{matrix} 
		\ \ \  0 ,\ (j,i)\notin\mathcal{E} \\  
		(\ref{1b}) ,(j,i)\in\mathcal{E}
	\end{matrix}\right .
\end{equation}
Define clipping function $[y]_{\underline{y}}^{\overline{y}}=\begin{cases}
	\overline y,& \text{ if } \overline y<y \\
	y,& \text{ if } \underline y\le y\le \overline y\\
	\underline y,& \text{ if } y<\underline y
\end{cases}$. 
Assume synaptic strengths are bounded with fixed sign:
$\underline{a}^- < a_{ij} < \overline{a}^- < 0, \forall (j,i) \in \mathcal{E}^-$ and
$0 < \underline{a}^+ < a_{ij} < \overline{a}^+, \forall (j,i) \in \mathcal{E}^+$.
If $a_{ij}$ exceeds bounds, it is clipped to the boundary value. This reflects the neuroscience principle that synapses have natural bounds. If nodes $i$ and $j$ are disconnected initially, no direct connection forms during evolution. Thus the graph topology is static, with only edge weights changing. Research in \cite{Roy2016,Ryan2015} shows lost synaptic strengths can be restored via neuromodulation even after amnesia, implying reversible yet bounded plasticity. Our model incorporates these features, assuming accessible yet constrained synaptic dynamics.

\begin{figure}[h]
	\begin{center}
		\includegraphics[scale=0.6]{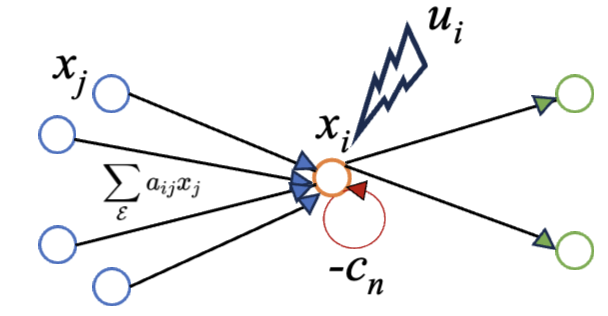}
	\end{center}
	\caption{ A diagram of the coupled model that describes the activity of neuron $x_i$.}
	\label{fig:model}
\end{figure}

In vector form, the model becomes
\begin{subequations}
	\label{eq:model}
	\begin{align}
		\dot x(t)&=-C_nx(t)+A(t)x(t)+ Bu,\ \  x\in\mathbb{R}^N,\label{2a}\\
		A(t+\tau)&=\left[C_a\circ A(t)+\phi(x(t)x(t)^T)\right]_{\underline A}^{\overline A},\label{2b}
	\end{align}
\end{subequations}
where $ A=[a_{ij}]\in\mathbb{R}^{N\times N}$ is the weight matrix with all its diagonal entries equal to zero and $u=[u_i]^T$ is the external stimulus vector. Let $``\circ"$ denote Hadamard product and $``\preceq "$ component-wise comparisons of matrices. $\underline{A}$ has zero entries when $(j,i)\notin\mathcal{E}$, $\underline{a_{ij}}^-$ if $(j,i)\in\mathcal{E}^-$, and $\underline{a_{ij}}^+$ if $(j,i)\in\mathcal{E}^+$. Similarly, $\overline{A}$ is defined with bounds $\overline{a}_{ij}^-, \overline{a}_{ij}^+$. Therefore, $\underline A\preceq  A\preceq \overline A$. $B=[b_{k_1}e_{k_1};\mathbf{0};b_{k_2}e_{k_2};\mathbf{0};\dots;b_{k_K}e_{k_K};\mathbf{0}]$ with $b_k$ a scalar weight and $e_k$ the $k$-th canonical vector of dimension $N$ determining the driver nodes $\{k_1,k_2,\dots,k_K\}$ of the graph formed by this system. In this paper, we mark $k_1$ is node $1$. $C_n$ is the diagonal matrices with $c_n>0$ as its entries. $C_a$ is formed by the corresponding $c_a^+$ and $c_a^-$, and $0<c_a^+,c_a^-<1$ exhibiting the synaptic depression property. It is noted that even though the terms $C_nx$ and $A(t)x$ in (\ref{2a}) can be combined, they have different meanings: the positive definite diagonal matrix $C_n$ denotes the decay rate and $C_nx$ represents the intrinsic dynamics of neurons, while $A$ denotes the connectome of the network that can be changing over time. This coupled model is illustrated in Fig. \ref{fig:model}.

\section{Dynamic Properties of the Model}
In this section, we analyze three properties of this dynamic neural network, which exhibits key features of neural systems.
\subsection{Boundedness}
From the view of biology, it is reasonable to assume the external stimuli and the activation function among neurons are bounded. Thus, in our network model, we have 
\begin{equation}
	\label{amp1}
	|u_i|\le u_{max},\  |\phi(\cdot)|\le \phi_{max}.
	\end{equation}
In practice, a commonly used activation function is the sigmoid \cite{Gerstner2014} and in this paper, we take the range of the sigmoid function to be $(-1,1)$ so that $\sup \phi_{max}=1$ and $\phi(0)=0$. From the dynamic equations (\ref{eq:model}), we define $\mathcal{X}=\{|x_i|\le x_{max}\}$, where $x_{max}=\frac{||Bu||_{\infty}}{c_n-\max\{d_{max}\overline{a}^+,d_{max}|\underline{a}^-|\}}$ denotes the maximum neural state and $|| \cdot||_\infty$ is the infinity norm. In neural network models, neuron states are used to represent the membrane potential of biological neurons. The membrane potential indicates the level of electrical polarization of the neuron's cell membrane, which is related to the neuron firing an action potential. Then, $x_{max}$ can be viewed as the maximum membrane potential. Since $B$ is a constant matrix and $u$ is bounded by $u_{max}$, $||Bu||_{\infty}$ is also bounded.

\begin{lemma}
	\label{lemma1}
	When assumptions (\ref{amp1}) hold and $\max\{d_{max}\overline{a}^+,d_{max}|\underline{a}^-|\}<c_n$, the trajectory of the coupled dynamics system (\ref{eq:model}) from any starting points $x_0\in\mathcal{X}$ will remain in $\mathcal{X}$.
	\end{lemma}

\begin{proof}
	(Sketch) The lemma asserts that $\mathcal{X}$ is an invariant set. Consider a solution $x^s$ of $(\ref{2a})$ starting in $\mathcal{X}$ and $|x_i|$ is the largest entry of $x^s$, and we assume $x_i=x_{max}=\frac{||Bu||_{\infty}}{c_n-\max\{d_{max}\overline{a}^+,d_{max}|\underline{a}^-|\}}$. The right-hand side of $(\ref{1a})$ becomes $-c_nx_i+\sum_{(j,i)} a_{ij}x_j+b_iu_i\le (-c_n+\sum_{(j,i)} a_{ij})x_i+b_iu_i\le (-c_n+\max\{d_{max}\overline{a}^+,d_{max}|\underline{a}^-|\})x_i+b_iu_i=-||Bu||_{\infty}+b_iu_i\le 0$, which derives the $x_i$ into the interior or at least stay on the boundary of the set $\mathcal{X}$. It is the same when $x_i$ approaches $-x_{max}$. Therefore, $\mathcal{X}$ is an invariant set. In addition, if a solution of equation $(\ref{2a})$ having initial conditions outside $\mathcal{X}$, say $|x_i|>x_{max}=\frac{||Bu||_{\infty}}{c_n-\max\{d_{max}\overline{a}^+,d_{max}|\underline{a}^-|\}}$, where $|x_i|$ is the largest entry of $x^s$, and without loss of generality, we assume $x_i>0$. When $\max\{d_{max}\overline{a}^+,d_{max}|\underline{a}^-|\}<c_n$, the right-hand side of $(\ref{1a})$ becomes $-c_nx_i+\sum_{(j,i)} a_{ij}x_j+b_iu_i\le (-c_n+\sum_{(j,i)} a_{ij})x_i+b_iu_i\le (-c_n+\max\{d_{max}\overline{a}^+,d_{max}|\underline{a}^-|\})x_i+b_iu_i<-||Bu||_{\infty}+b_iu_i\le 0$ until $x_i=x_{max}$. It means the solution will decrease until it evolves into the set $\mathcal{X}$. Therefore, we can conclude that $\mathcal{X}$ is an invariant forward and attractive set. A similar rigorous proof is given in \cite{menara2019} using comparison lemma \cite[pp. 102-103]{Khalil2022}.
\end{proof}

The boundedness property holds biological justification, as neural interconnectivity magnitudes must remain within constrained ranges, and states of neurons given in terms of membrane potentials exhibit saturation given extreme input values.

\subsection{Stability}
It is observed that when an organism is in a resting state, neurons are not actively engaged in electrical communication, which is referred to as having the membrane potential in a steady state. After experiencing an external stimulus, the neurons' membrane potential changes in response to the stimulus due to the charged ions crossing the cell membrane, but after a period of time, the membrane potential re-stabilizes \cite{Perkel1978}. Our proposed system (\ref{eq:model}) also exhibits this feature.
\begin{theorem}
	\label{thm1}
	When conditions in lemma \ref{lemma1} are satisfied, the neuromimetic dynamic model (\ref{2a}) is asymptotically stable without external stimuli and the whole system ($\ref{eq:model}$) is stable.
	\end{theorem}
\begin{proof}
	Denote the $p-$th time slot by $(p\tau,(p+1)\tau],p=0,1,2,\dots$, and let $t_p\in(p\tau,(p+1)\tau]$ be an arbitrary time instant in the time slot. $A_p$ is the corresponding weight matrix during this period, which is constant according to (\ref{2b}). Therefore, $A(t)$ is a piecewise constant matrix. Consider any time slot and denote the state matrix $H_p=-C_n+A_p$ in (\ref{2a}), where $C_n$ is a diagonal matrix. Since $\sum_{j=1,j\neq i}^{N}a_{ij}\le \max\{d_{max}\overline{a}^+,d_{max}|\underline{a}^-|\}$, when the condition in lemma \ref{lemma1} is satisfied, for each row $i$ of $-H_p$, $-H_p(i,i)=c_n>\max\{d_{max}\overline{a}^+,d_{max}|\underline{a}^-|\}\ge\sum_{j=1,j\neq i}^{N}-a_{ij}=\sum_{j=1,j\neq i}^{N}-H_p(i,j)$. Therefore, $-H_p$ is a Hermitian diagonally dominant matrix. According to Gershgorin circle theorem, we obtain $-H_p$ is positive definite and $H_p$ is Hurwitz for all $p$. Thus, the close-form solution of the state in time slot $(p\tau,(p+1)\tau]$ can be expressed as
\begin{equation}
	\begin{split}
x(t_p)&=e^{H_p(t_p-p\tau)}e^{H_{p-1}\tau}\dots e^{H_0\tau}x_0,\\
\end{split}
\end{equation}
whose trajectory is continuous. Since $H_p$ is Hurwitz and $t_p-p\tau>0$, $|e^{H_p(t_p-p\tau)}|\le\kappa e^{-\lambda(t_p-p\tau)}=\kappa' e^{-\lambda t_p}$, where $\lambda$ can be $|\lambda_{max}(H_p)|$ and $\kappa'=\kappa e^{\lambda p\tau}$. $\lambda_{max}(H_p)$ is the maximum eigenvalue of $H_p$. Term $e^{H_{p-1}\tau}\dots e^{H_0\tau}$ is a constant matrix, then $|e^{H_{p-1}\tau}\dots e^{H_0\tau}|\le\alpha e^{-\lambda'\tau}$, where $\lambda'$ can take $|\lambda_{max}(H_{p-1}+\dots+ H_0)|$ and it can be merged into the term $\kappa'$ to be $\hat\kappa$. Therefore, we can write $|e^{H_p(t_p-p\tau)}e^{H_{p-1}\tau}\dots e^{H_0\tau}|\le \hat\kappa e^{-\lambda t_p}$, which implies (\ref{2a}) is asymptotically stable. Meanwhile, $A$ is bounded naturally and thus, the whole system ($\ref{eq:model}$) is stable.
\end{proof}
\begin{corollary}
	\label{c1}
	There is a unique equilibrium $(x,A)=(x^{eq},A^{eq})$ of the coupled system ($\ref{eq:model}$) without external stimuli, where $x^{eq}=\mathbf{0}$, $A^{eq}_{ij}=\underline{a}^+$ if $(j,i)\in\mathcal{E}^+$, $A^{eq}_{ij}=\overline{a}^-$ if $(j,i)\in\mathcal{E}^-$ and others are zero.
	\end{corollary}
\begin{proof}
	It can be easily observed that $(x^{eq},A^{eq})$ satisfy the equations $0=-C_nx^{eq}+A^{eq}x^{eq}$ and $A^{eq}=\left[C_a\circ A^{eq}+\phi(x^{eq}(x^{eq})^T)\right]_{\underline A}^{\overline A}$, so that it is an equilibrium of the system ($\ref{eq:model}$). Meanhwile, from Theorem \ref{thm1}, we obtain the state matrix of (\ref{2a}) is Hurwitz and the neural dynamic is asymptotically stable, so that $x^{eq}$ is unique. With $t\rightarrow\infty$, $x\rightarrow\mathbf{0}$ and $\phi(xx^T)=\mathbf{0}$. Since $C_a$ has all its entries in $(0,1)$, when $t\rightarrow\infty$, $|A(t+\tau)|<|A(t)|$ until reaching the boundary. Therefore, $a_{ij}$ asymptotically converges to $\underline{a}^+$ if $(j,i)\in\mathcal{E}^+$ and $\overline{a}^-$ if $(j,i)\in\mathcal{E}^-$ without external stimuli, which implies that the coupled system $(\ref{eq:model}$) has a unique equilibrium $(x^{eq},A^{eq})$ when there is no input.
	\end{proof}

\subsection{Structural Controllability}
In classical control theory, controllability of the linear time-invariant system $\dot{x} = \hat{A}x + \hat{B}u$ is determined by the controllability matrix $\hat{C} = [\hat{B},\hat{A}\hat{B},\dots,\hat{A}^{N-1}\hat{B}]$ having full rank. By analogy with control theory, the graph is said to be a \emph{controllable graph} if and only if the controllability matrix derived from the adjacency (weight) matrix and input matrix is full rank \cite{Kailath1980}. However, this rank condition is impractical since the matrices $\hat{A}$ and $\hat{B}$ are often inaccessible, with measurements also imprecise. Thus, \cite{lin1974} proposed a system is structurally controllable if almost all nonzero weight combinations of $\hat{A}$ and $\hat{B}$ yield a full-rank $\hat{C}$. Under this definition, \cite{lin1974} showed linear systems with a single control node are structurally controllable if the graph $G$ has no inaccessible nodes and no dilations, implying cactus-spanned graphs confer structural controllability.
To pursue an alternative approach, we say that a connected graph $\mathcal{G}_a=(\mathcal{V}_a,\mathcal{E}_a)$ is spanned by a connected graph $\mathcal{G}_b=(\mathcal{V}_b,\mathcal{E}_b)$ when $\mathcal{V}_a=\mathcal{V}_b$ and $\mathcal{E}_b\subseteq \mathcal{E}_a$. 

 In \cite{Mayeda1981,Kang1996}, authors extended \cite{lin1974} to the multi-input case and concluded that the graph is structurally controllable if and only if there exist cacti that span the graph. In \cite{menara2019}, graph symmetry was studied and it was shown that a symmetric graph control system with a single control node is structurally controllable if and only if the graph without symmetry is structurally controllable. However, all the above work focuses on LTI systems, while controllability proofs for hybrid systems like (\ref{eq:model}) remain challenging. Here, we give the definition of structural controllability for a linear time-varying system.


\begin{definition}
	\label{definition1}
	(Structural Controllability) \cite{Hou2016} The linear time-varying switching network $\mathcal{G}$ is structurally controllable if there exists a controllable linear temporally switching system having the same structure with $\mathcal{G}$.
\end{definition}

\begin{lemma}
	\label{lemma2}
	(\ref{eq:model}) is structurally controllable if there exists one controllable graph that has the same structure as $A$.
	\end{lemma}
\begin{proof}
	Notice that our proposed model (\ref{eq:model}) has a specific structure that, in each time slot, (\ref{2a}) is a linear time-invariant system since $A$ is constant. Meanwhile, in different time slots, even if matrices $A$ are different, they share the same structure -- the location of zero entries is fixed over time, while the non-zero weights are allowed to vary. From definition \ref{definition1} of the structural controllability for the time-varying system, we reach the conclusion of lemma \ref{lemma2}.
	\end{proof}

The general graph theoretic definition of a cactus graph can be specialized without modification to the case of symmetric digraphs. In what follows, we construct sym-cactus structures by means of sym-cycles and bi-directional connections with symmetric weights. 
\begin{definition}\label{definiton2}
	(Sym-cactus) \cite{menara2019} A sym-cactus is a connected digraph $\mathcal{G}$ with symmetric weights and defined as $\mathcal{G}(\mathcal{V},\mathcal{E})={\textstyle \overline{\bigcup}_{i=1}^{m}} \mathcal{G}_i(\mathcal{V}_i,\mathcal{E}_i)$, where $\mathcal{G}_i$ is symmetric cycle or a single node and $\mathcal{V}_i\cap\mathcal{V}_j=\emptyset, \forall i\neq j$. 
	\end{definition}
\begin{figure}[h]
	\begin{center}
		\includegraphics[scale=0.56]{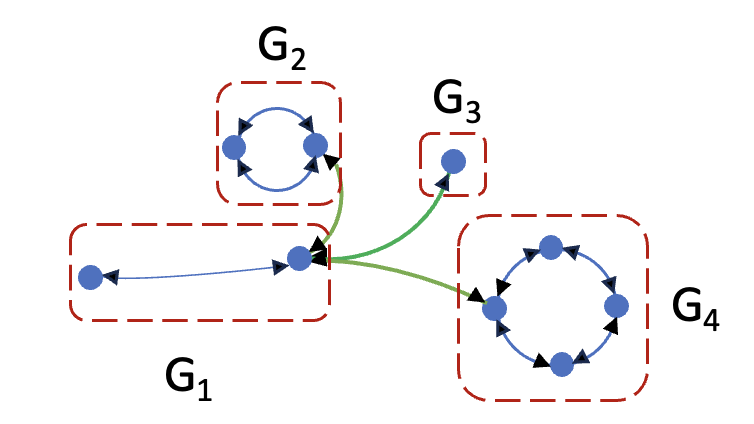}
	\end{center}
	\caption{The diagram shows a sym-cactus graph $\mathcal{G}$ with $m=4$ subgraphs in red dotted boxes. The green lines represent connecting edges of different subgraphs.}
	\label{fig:cup}
\end{figure}
Here, $\mathcal{G}(\mathcal{V},\mathcal{E})=\overline{\bigcup}_{i=1}^{m}\mathcal{G}_i$ represents $\mathcal{V}={\bigcup_{i=1}^{m}}\mathcal{V}_i$, $\mathcal{E}={\bigcup_{i=1}^{m}}\mathcal{E}_i\cup\overline{\mathcal{E}}$. $\overline{\mathcal{E}}$ is the set of edge pairs that satisfies for all $i\in\{2,3,\dots,m\}$, there is a unique edge pair $(d_1^{i},d_2^{i})\in \overline{\mathcal{E}}$,where $d_1^{i}\in\mathcal{V}_i$, $d_2^{i}\in{\bigcup}_{j=1}^{i-1}\mathcal{V}_j$ and $(d_2^{i},d_1^{i})\in \overline{\mathcal{E}}$. For example, in Fig. \ref{fig:cup}, the subgraphs $G_1, G_2,G_3$ and $G_4$ are depicted in red dotted boxes and green lines represent the symmetric edges in the set $\overline{\mathcal{E}}$ that connect vertices between different subgraphs. 

\begin{definition}
	\label{definition3}
	(Generalized Sym-cactus Structure) A symmetric digraph is said to have a generalized sym-cactus structure if it is spanned by a disjoint union of sym-cacti and may have connections between them.
	\end{definition}

\begin{figure}[h]
	\begin{center}
		\includegraphics[scale=0.5]{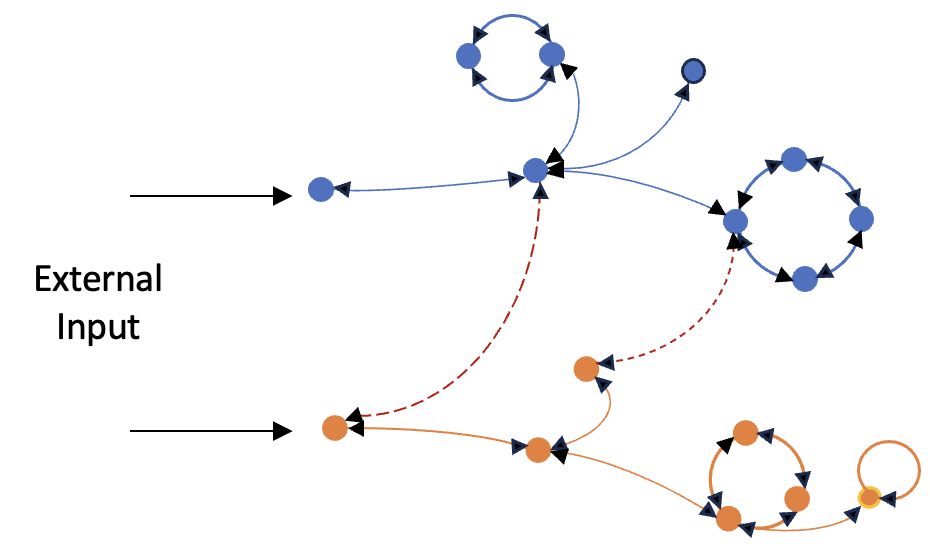}
	\end{center}
	\caption{ A diagram of a generalized sym-cactus structure formed by two sym-cacti in blue and orange color. Red dotted lines denotes connections between nodes from two different cacti.}
	\label{fig:cacti}
\end{figure}

\begin{theorem}
	The symmetric weighted digraph $\mathcal{G}$ formed by dynamic system (\ref{eq:model}) is structurally controllable with multiple control nodes denoted by $\{k_1,k_2,\dots,k_K\}$ if the digraph has a generalized sym-cactus structure rooted from $\{k_1,k_2,\dots,k_K\}$.
\end{theorem}

\begin{proof}
	From lemma \ref{lemma2}, we know that to prove the digraph $\mathcal{G}$ of model (\ref{eq:model}) is structurally controllable is the same as to find a controllable graph with weight matrix $A$. Following the result in \cite[Theorem 3.3]{menara2019} that a symmetric weights digraph with a single control node is structurally controllable with symmetric weights if it is spanned by a sym-cactus rooted at the control node, we obtain for each subgraph $\mathcal{G}_{k'}$ with $k'\in\{1,2,\dots,K\}$, that it is structurally controllable, which means that we can find a choice of weights to form a weight matrix $A_{k'}$ and corresponding $B_{k'}$ such that $\mathcal{G}_{k'}$ is controllable. If we combine all these disjoint subgraphs, the weight matrix of the whole graph $\mathcal{G'}$ can be written as $A'=\begin{bmatrix}
		A_1& \mathbf{0} &\cdots  & \mathbf{0}\\
		\mathbf{0}& A_2 &\mathbf{0}  &\mathbf{0} \\
		\vdots& \vdots &\ddots   &\vdots \\
		\mathbf{0}& \mathbf{0} &\cdots  &A_K
	\end{bmatrix}$ with $B=[b_1e_1;\mathbf{0};\dots;b_ke_k;\mathbf{0};\dots;b_Ke_K;\mathbf{0}]$. Given that $(A_{k'},B_{k'})$ is a controllable pair for all $k'$, the controllability matrix $C_{k'}$ is full rank. Therefore, the controllability matrix $C'$ formed by $(A',B)$ has $N$ linearly independent rows, which is full rank. In addition, if more edges are added and weights are randomly chosen, controllability remains a generic property. Here is an example.
\begin{example}
	Suppose there are two controllable subgraphs $A_1=\begin{bmatrix}
		0& 2\\
		2&0
	\end{bmatrix}$ and $A_2=\begin{bmatrix}
	0&1  &2 \\
	1& 0 &1 \\
	2& 1 &0
\end{bmatrix}$ with the first node in these two subgraphs being the driver node. Then combining them together, 
	$A'=\begin{bmatrix}
		0& 2& 0 &0 &0\\
		2&0 & 0 &0 &0\\
		0& 0& 0&1  &2 \\
		0& 0& 1& 0 &1 \\
		0& 0& 2& 1 &0
	\end{bmatrix}$, $B=\begin{bmatrix}
	1&  0&  0& 0 &0 \\
	0&  0&  0&  0&0 \\
	0& 0 &  1&  0&0 \\
	0&  0&  0&  0&0\\
	0&  0&  0&  0&0
\end{bmatrix}$, where the corresponding controllability matrix $C'=(B\ A'B\  A'^2B\ A'^3B\ A'^4B)$ is full rank. If at each node we add a self-loop and a connection between node $2$ and $4$ is added as shown in Figure \ref{fig:ex}(b), after choosing the weight, the weight matrix of the new graph becomes $A_{new}=\begin{bmatrix}
-3& 2& 0 &0 &0\\
2&-4& 0 &-0.5 &0\\
0& 0& -4&1  &2 \\
0& -0.5& 1& -3 &1 \\
0& 0& 2& 1 &-4
\end{bmatrix}$ with $B$ unchanged. The corresponding controllability matrix $C_{new}=(B\ A_{new}B\  A_{new}^2B\ A_{new}^3B\ A_{new}^4B)$ is still full rank.
	\end{example}
\begin{figure}[h]
	\begin{center}
		\includegraphics[scale=0.38]{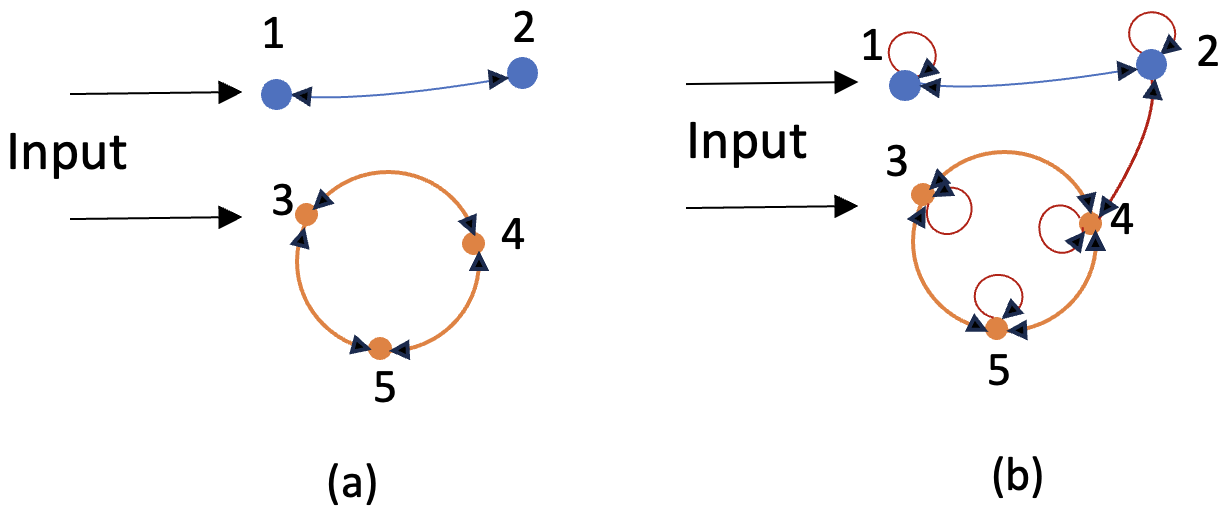}
	\end{center}
	\caption{ (a) is the graph with weight matrix $A'$, which contains two subgraphs, while (b) is the new graph by adding self loops and an edge between node 2 and 4.}
	\label{fig:ex}
\end{figure}
Thus, according to definition \ref{definition3}, we have a generalized sym-cactus structure that is structurally controllable.
	\end{proof}
\begin{remark}
	(Grow and prune connections) While our model focuses on synaptic plasticity rather than growing novel connections or pruning lesser utilized synapses, remodeling is crucial developmentally and pathologically. Per Theorem \ref{thm1}, the system maintains structural controllability despite removing or adding connections, contingent on preserving the generalized sym-cactus topology. Moreover, the system (\ref{eq:model}) guarantees stability, given dominant cellular decay rates ($c_n$) amidst such structural changes (i.e. $c_n>\max\{d_{max}\overline{a}^+,d_{max}|\underline{a}^-|\}$). Although not directly incorporated, both controllability and stability theoretically allow for graph evolution through growth and atrophy, provided key topological and dynamical constraints remain satisfied.
	\end{remark}
\section{Simulations}
\begin{figure}[h]
	\begin{center}
		\includegraphics[scale=0.5]{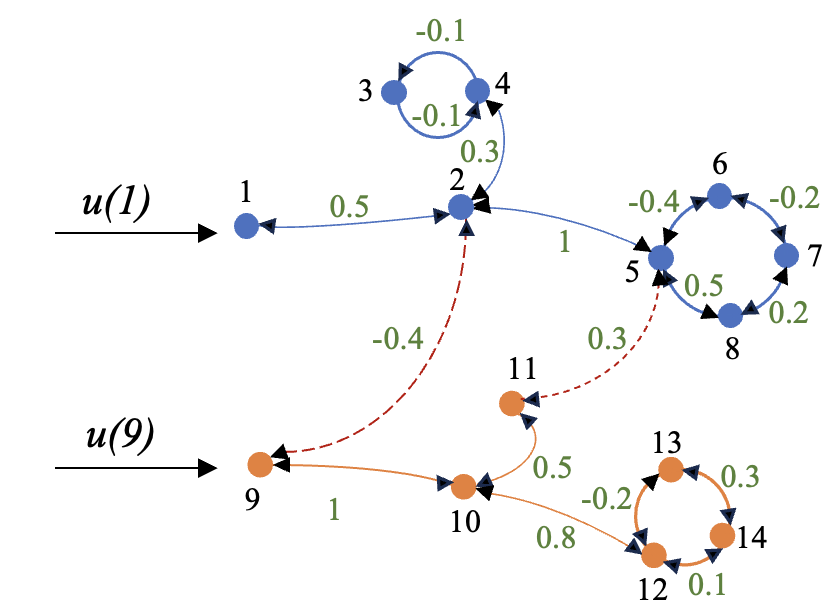}
	\end{center}
	\caption{The figure shows a generalized sym-cactus graph with 14 nodes. Green numbers denote edge weights at initial conditions. Each node also has a self-loop, but not exhibited here. Only nodes 1 and 9 have external input.}
	\label{fig:sym_weight}
\end{figure}

This section provides simulations validating the theorems from previous sections. We implement the model (\ref{eq:model}) using a 14-neuron generalized sym-cactus topology, with edge weights as in Fig. \ref{fig:sym_weight} representing the weight matrix $A$. Decay terms at each node act as self-loops, thus not affecting the cacti structure and not exhibiting in the figure. From the network in Fig. \ref{fig:sym_weight}, excluding self-loops, the maximum out-degree is $d_{max}=4$.
 There are two control nodes $k_1=1$ and $k_2=9$, so that $B=[e_1;\mathbf{0}^{7\times14};e_{9};\mathbf{0}^{5\times14}]$ with $b_1=b_{9}=1$ and $e_1,e_{9}$ dimension $14$. Meanwhile, we set $\underline{a}^-=-1,\overline{a}^-=-0.05$ and $\underline{a}^+=0.05,\overline{a}^+=1$. The decay rate of each neuron is set to be $-c_n=-4.1$, which satisfies the condition $c_n-\max\{d_{max}\overline{a}^+,d_{max},\underline{a}^-\}>0$. It is worth noting that $(A,B)$ pair is controllable at the beginning. The time interval of updating $A$ is $\tau=0.2$. 

\begin{figure}[h]
	\begin{center}
		\includegraphics[scale=0.26]{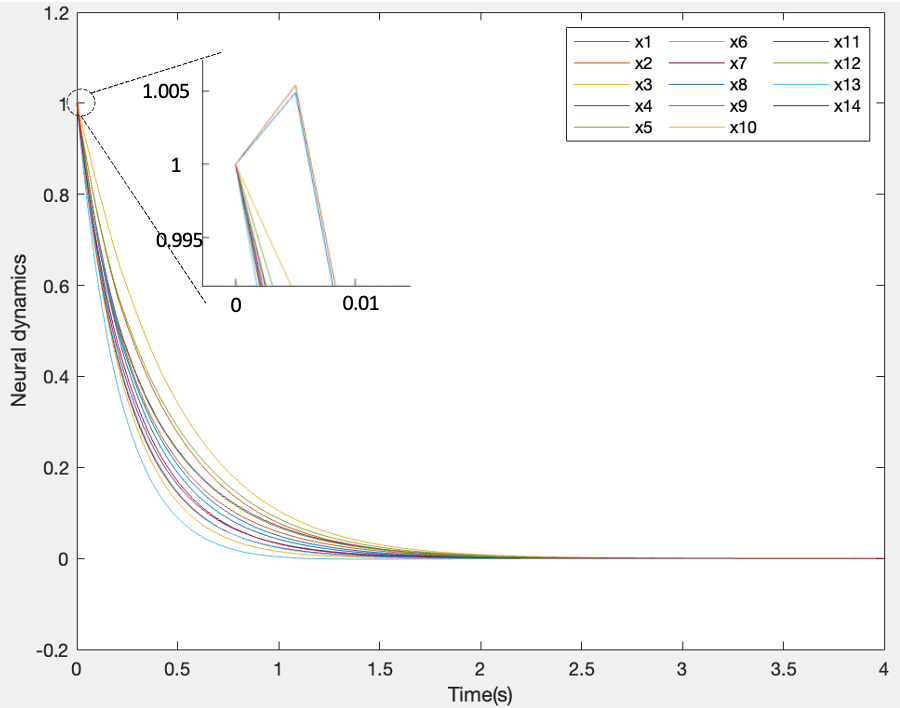}
		\includegraphics[scale=0.242]{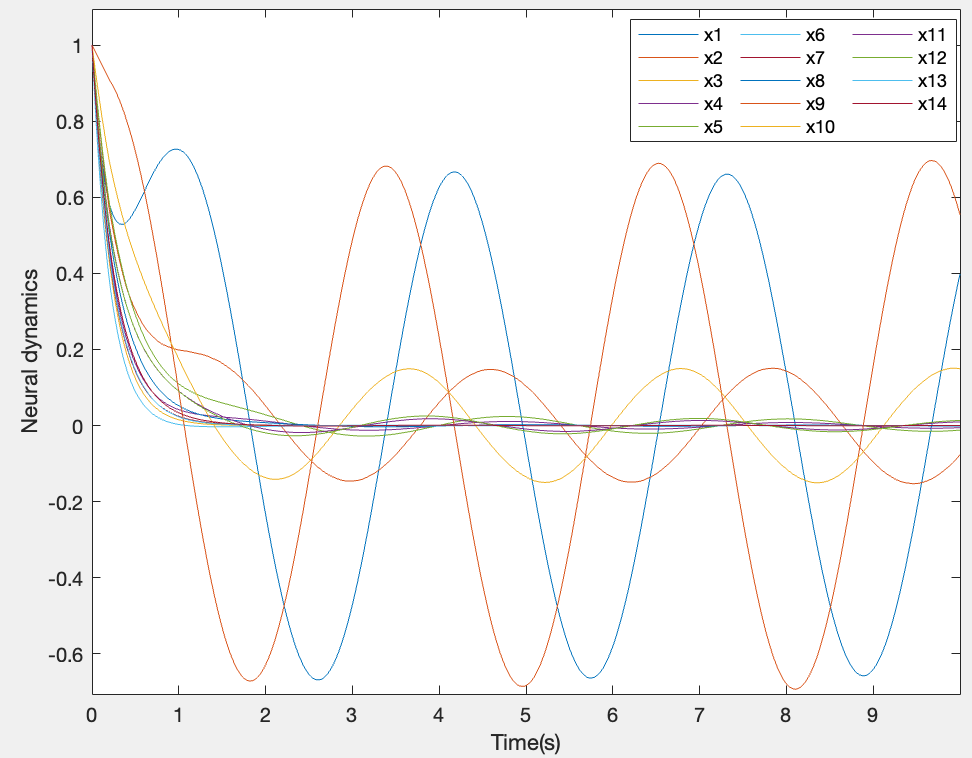}
	\end{center}
	\caption{Figure (a) is an example of the network in Fig. \ref{fig:sym_weight} with external input on node 1 and 9 to be impulse function $\delta(0)=2$, while (b) is using the input to be $3\sin(2t)$ and $3\cos(2t)$.}
	\label{fig:state}
\end{figure}

Suppose all neural states are $1$ at the beginning. We set the external input on node $1$ and $9$ to be an impulse function $\delta(t)=\begin{cases}
	2,&  \text{t\ =\ 0} \\
	0,& \text{otherwise}
\end{cases}$ that is $u(1)=2$ and $u(9)=2$. $x_{max}$ is 20. Another example is using the input $u(1)=3\sin(2t)$ and $u(9)=3\cos(2t)$. $x_{max}$ is 10 at this time. Results are shown in Fig. \ref{fig:state}. It can be observed that both neural dynamics are bounded within the $\mathcal{X}$ identified in lemma \ref{lemma1} and the impulse response shown in Fig. \ref{fig:state} converges to the origin, which is compatible with theorem \ref{thm1}.

\begin{figure}[h]
	\begin{center}
		\includegraphics[scale=0.148]{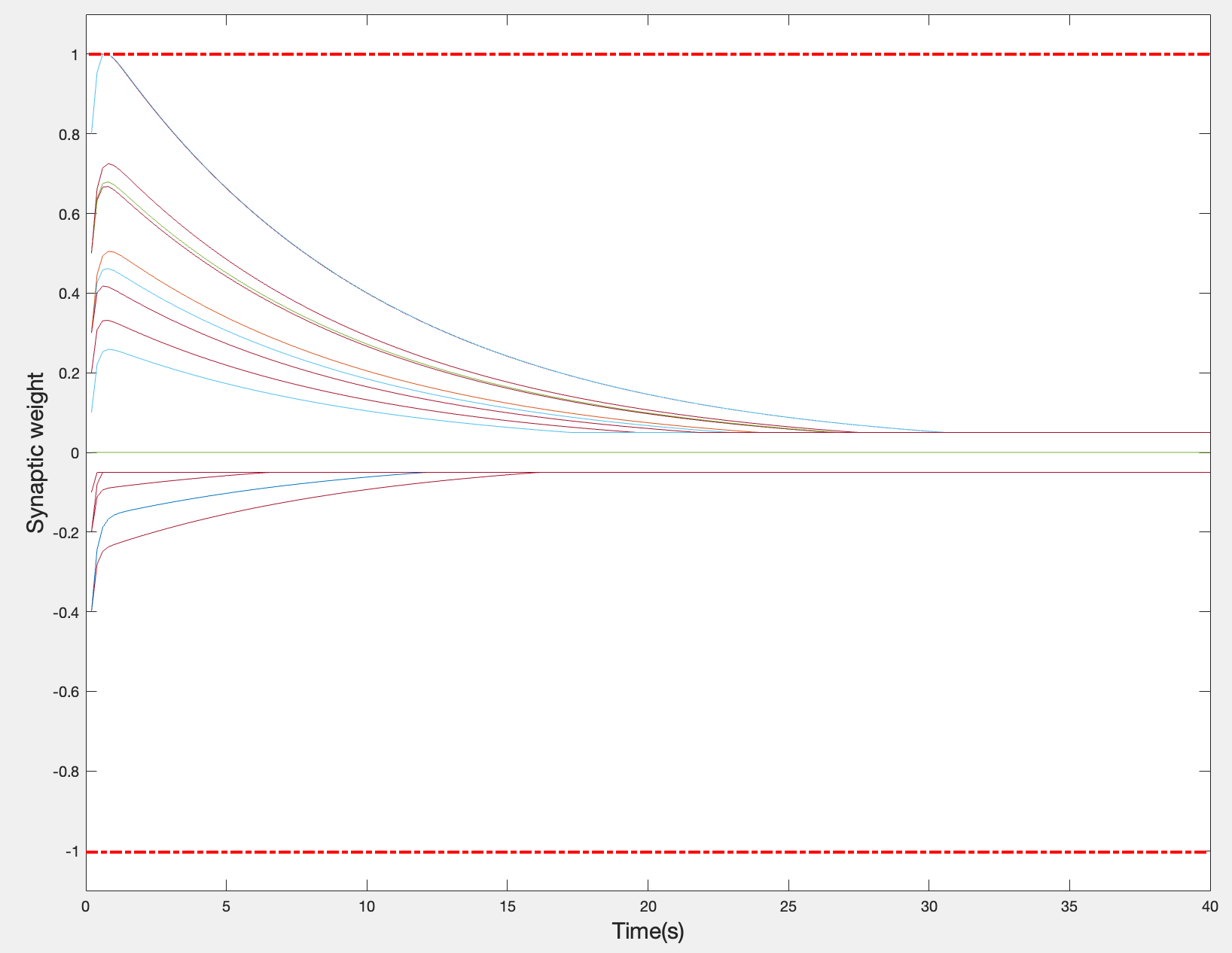}
		\includegraphics[scale=0.242]{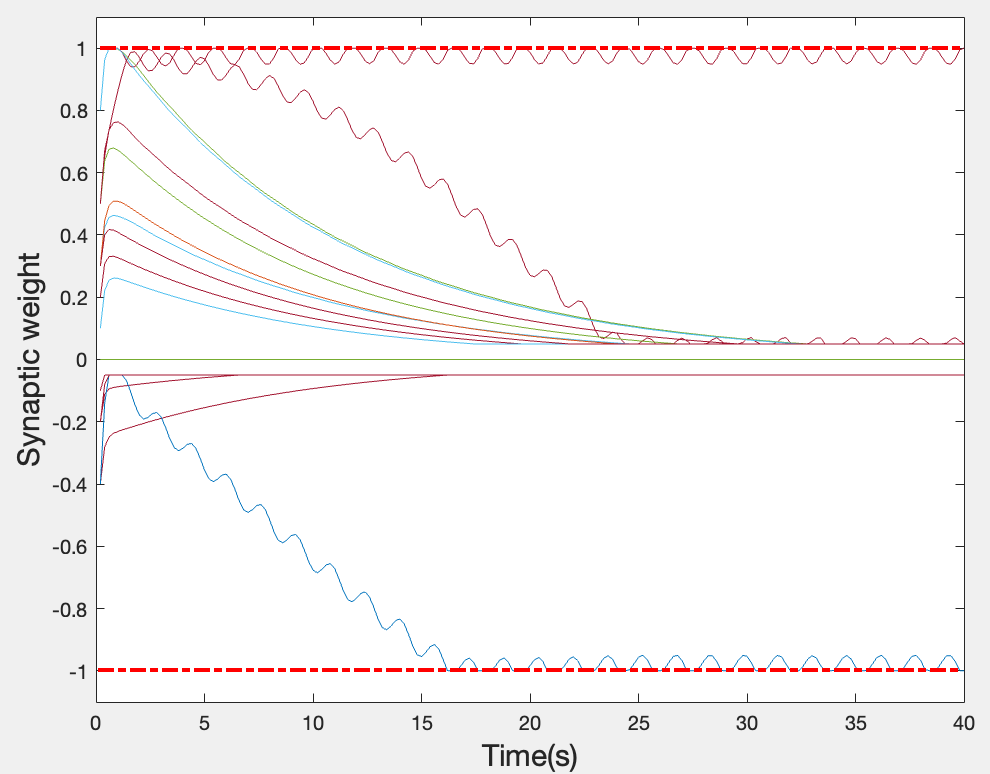}
	\end{center}
	\caption{Synaptic weights over time (excluding self-loop). The red dotted lines represent the bounded value $\underline{a}^-$ and $\overline{a}^+$.}
	\label{fig:sim_weight}
\end{figure}

Furthermore, Fig. \ref{fig:sim_weight} depicts the evolution of synaptic weights between neurons over 40 seconds under the two input cases corresponding to Fig. \ref{fig:state}. The weights evolve within the bounds [$\underline{a}^-$, $\overline{a}^+$], with the distinct inputs producing different $A$ matrices. With the impulse input, when external stimulation ceases, the weights converge to $\underline{a}^+$ or $\overline{a}^-$ based on initial sign, validating Corollary \ref{c1} as shown in Fig. \ref{fig:state} (a) and Fig. \ref{fig:sim_weight} (a). Overall, the simulations align with the theoretical weight boundedness and stability results.

\begin{figure}[h]
	\begin{center}
		\includegraphics[scale=0.227]{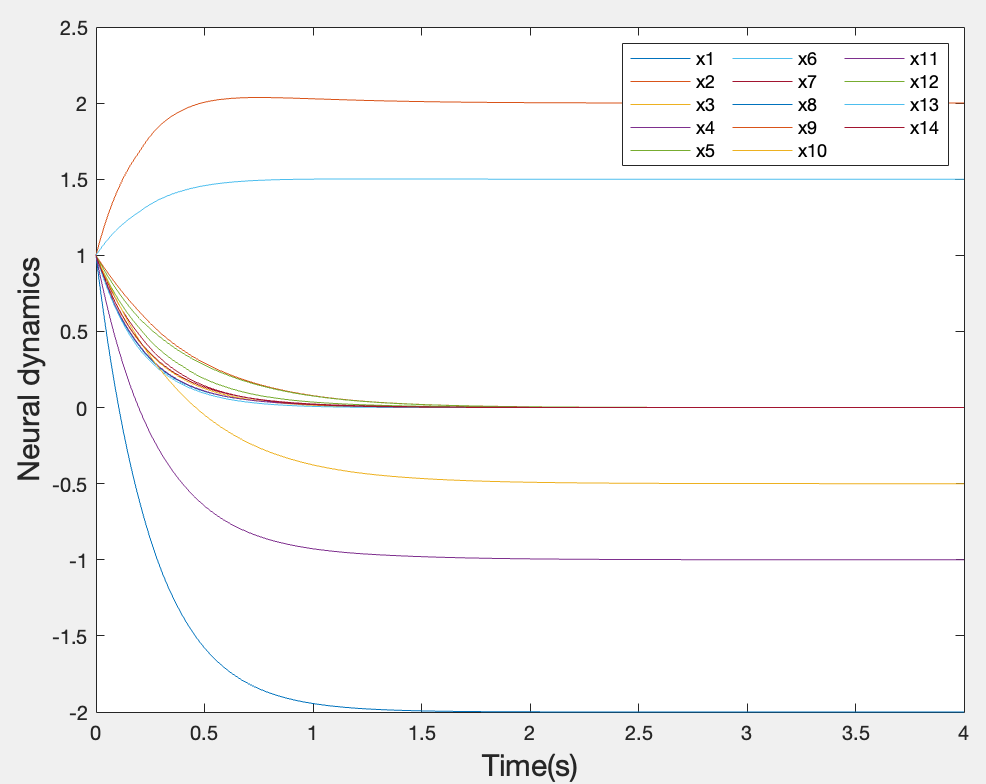}
		\includegraphics[scale=0.227]{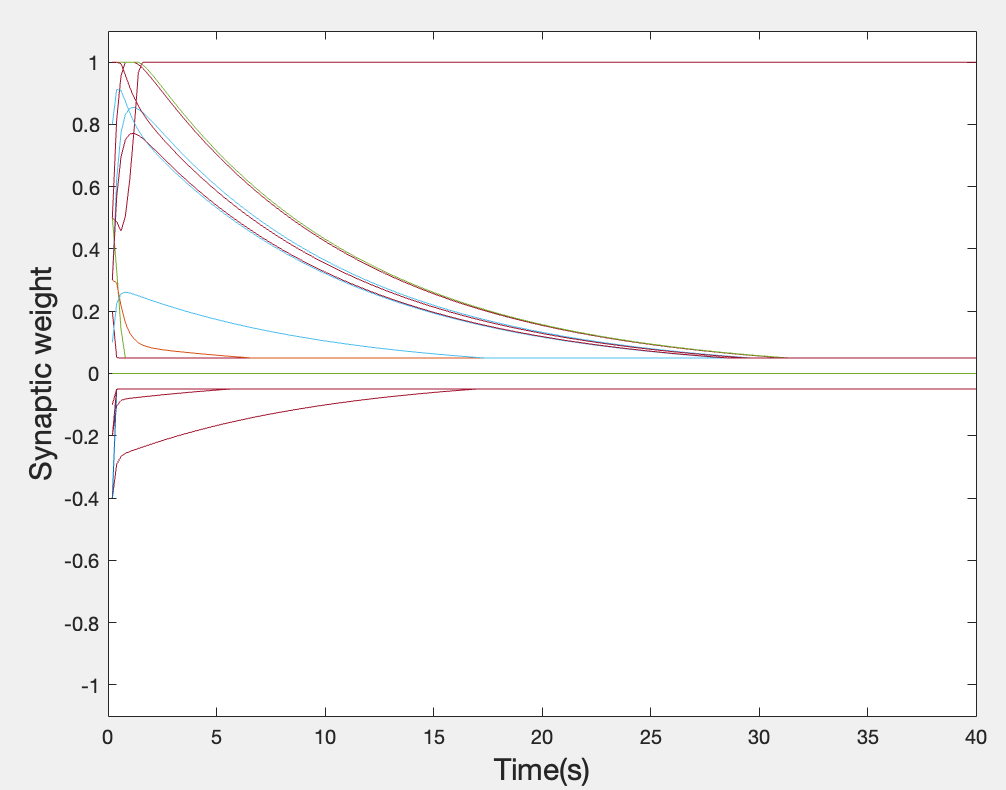}
	\end{center}
	\caption{(a) 
	Simulation of the model (\ref{eq:model}) using Fig. \ref{fig:sym_weight} topology reaches desired states under optimal control solved via Riccati equation. (b) Synaptic weight dynamics over 40 seconds.}
	\label{fig:optimalcontrol}
\end{figure}

Additionally, the controllability matrix rank remains full throughout evolution, affirming the generic controllability property. Simulations demonstrate driving the initial Fig. \ref{fig:sym_weight} network to desired states: $x_2=2, x_8=-2, x_{10}=-0.5, x_{13}=1.5, x_{11}=-1$ with other states zero. The input is feedback control $u=-Kx$ where $K$ comes from the LQR algorithm solving the algebraic Riccati equation. The cost function uses identity matrix $Q$ and $R=1$. Fig. \ref{fig:optimalcontrol} (a) shows the neuron dynamics, verifying that this feedback law steers the a generalized sym-cactus system to the specified states. Fig. \ref{fig:optimalcontrol} (b) displays the synaptic weight evolution. Together, the simulations align with and demonstrate the theoretical controllability results.

\section{Conclusion and future work}

In this work, we have reported a neuromimetic dynamic model that captures neural features like synaptic weights that evolve according to Hebbian learning principles with symmetric weights and multiple control nodes. Through theoretical proofs and simulations, it exhibits a boundedness feature when there are external inputs and stability without inputs. Since controllability is a generic property when considering the determinant of the controllability matrix, the model can finally converge to a controllable system by randomly choosing connection weights at the beginning. 

While the results exhibit important features of brain networks,we note there have been criticisms of Hebbian learning in that simultaneous neural excitation may not always indicate direct connectivity or causal influence. Furthermore, while some studies like \cite{menara2019} suggest symmetrical anatomical connectivity patterns, the symmetric weights assumption among brain regions is perhaps strong.  Future work will focus on designing a general asymmetric model that overcome these restrictions and matches the diffusion MRI data \cite{Gu2015}, which displays the brain's regional activity measured by functional magnetic resonance imaging. Broader applications such as interventions to influence brain dynamics will also be explored.





\bibliographystyle{IEEEtrans}

\begin{thebibliography}{3}
	
\bibitem{cook2019whole}
S.J. Cook, T.A. Jarrell, C.A. Brittin et al.\ ``Whole-animal connectomes of both Caenorhabditis elegans sexes'', {\em Nature} 571, 63–71 (2019). https://doi.org/10.1038/s41586-019-1352-7

\bibitem{mcculloch1943logical}
WS McCulloch, W. Pitts, ``A logical calculus of the ideas immanent in nervous activity'', {\em The bulletin of mathematical biophysics} 1943 Dec;5:115-33.

\bibitem{hodgkin1952currents}
AL. Hodgkin and AF. Huxley, ``Currents carried by sodium and potassiumions through the membrane of the giant axon of Loligo'', {\em The Journal of physiology}, 1952 Apr 4;116(4):449.

\bibitem{hopfield1982neural}
JJ. Hopfield, ``Neural networks and physical systems with emergent collective computational abilities'', {\em Proceedings of the national academy of sciences}, (PNAS), 1982 Apr;79(8):2554-8.

\bibitem{Sun2022}
Z. Sun and J. Baillieul. Neuromimetic Linear Systems—Resilience and Learning. In 2022 IEEE 61st Conference on Decision and Control (CDC) (pp. 7388-7394). IEEE, 2022.

\bibitem{Baillieul2019}
J. Baillieul, ”Perceptual Control with Large Feature and Ac- tuator Networks,” 2019 IEEE 58th Conference on Decision and Control (CDC), Nice, France, 2019, pp. 3819-3826, doi: 10.1109/CDC40024.2019.9029615.

\bibitem{Gaiteri2016}
C. Gaiteri, S. Mostafavi, C.J. Honey, P.L. De Jager and D.A. Bennett, Genetic variants in Alzheimer disease—molecular and brain network approaches. Nature Reviews Neurology, 12(7), pp.413-427, 2016.

\bibitem{Sun2023}
Z. Sun and J. Baillieul, Emulation Learning for Neuromimetic Systems. arXiv preprint arXiv:2305.03196, 2023.

\bibitem{Gerstner2014}
W. Gerstner, W. M. Kistler, R. Naud, and L. Paninski. Neuronal Dynamics: From Single Neurons To Networks and Models of Cognition. Cambridge University Press, 2014, ISBN 9781107635197.

\bibitem{Sherrill2015}
K.R. Sherrill, E.R. Chrastil, R.S. Ross, U.M. Erdem, M.E. Hasselmo and C.E. Stern. Functional connections between optic flow areas and navigationally responsive brain regions during goal-directed navigation. Neuroimage, 118, pp.386-396, 2015.

\bibitem{Veronica2022}
C. Veronica, F. Bullo, and G. Russo. "Contraction analysis of hopfield neural networks with hebbian learning." In 2022 IEEE 61st Conference on Decision and Control (CDC), pp. 622-627. IEEE, 2022.

\bibitem{Krotov2019}
D. Krotov and J. J. Hopfield. Unsupervised learning by competing hidden units. Proceedings of the National Academy of Sciences, 116:201820458, 03 2019. doi:10.1073/pnas.1820458116.

\bibitem{Citri2008}
A. Citri and R.C. Malenka, Synaptic plasticity: multiple forms, functions, and mechanisms. Neuropsychopharmacology, 33(1), pp.18-41, 2008.

\bibitem{Ryan2022}
T.J. Ryan and P.W. Frankland, 2022. Forgetting as a form of adaptive engram cell plasticity. Nature Reviews Neuroscience, 23(3), pp.173-186.

\bibitem{Kailath1980}
T. Kailath, Linear Systems. Englewood Cliffs, NJ, USA: Prentice-Hall, 1980.

\bibitem{Roy2016}
D.S. Roy, A. Arons, T.I. Mitchell, M. Pignatelli, T.J. Ryan and S. Tonegawa, Memory retrieval by activating engram cells in mouse models of early Alzheimer’s disease. Nature, 531(7595), pp.508-512, 2016.

\bibitem{Ryan2015}
T.J. Ryan, D.S. Roy, M. Pignatelli, A. Arons and S. Tonegawa. Engram cells retain memory under retrograde amnesia. Science, 348(6238), pp.1007-1013, 2015.

\bibitem{Perkel1978}
D.H. Perkel and B.R.I.A.N. Mulloney, Electrotonic properties of neurons: steady-state compartmental model. Journal of Neurophysiology, 41(3), pp.621-639, 1978.

\bibitem{Khalil2022}
H. K. Khalil. Nonlinear Systems. Prentice Hall, 3 edition, 2002, ISBN 0130673897.

\bibitem{Kang1996}
K. Li, Y. Xi, and Z. Zhang, Cactus-Like and Structural Controllability of Interconnected Dynamical System. IFAC Proceedings Volumes, 29(1), pp.4452-4457, 1996

\bibitem{lin1974}
C. Lin, "Structural controllability," in IEEE Transactions on Automatic Control, vol. 19, no. 3, pp. 201-208, June 1974, doi: 10.1109/TAC.1974.1100557.

\bibitem{menara2019}
T. Menara, D.S. Bassett, and F. Pasqualetti. Structural controllability of symmetric networks. IEEE Transactions on Automatic Control, 64(9), pp.3740-3747, 2018.

\bibitem{Mayeda1981}
H. Mayeda, “On structural controllability theorem,” IEEE Trans. Autom. Control, vol. 26, no. 3, pp. 795–798, Jun. 1981.

\bibitem{Hou2016}
B. Hou, X. Li, and G. Chen, Structural controllability of temporally switching networks. IEEE Transactions on Circuits and Systems I: Regular Papers, 63(10), pp.1771-1781, 2016.

\bibitem{Gu2015}
S. Gu et al., “Controllability of structural brain networks,” Nature Commun., vol. 6, 2015, Art. no. 8414.

\bibitem{Galan2008}
R. F. Galan, “On how network architecture determines the dominant patterns of spontaneous neural activity,” PLoS ONE, vol. 3, no. 5, 2008, Art. no. e2148.
\bibitem{Abb2000}
L.F. Abbott and S.B. Nelson, Synaptic plasticity: taming the beast. Nature neuroscience, 3(11), pp.1178-1183, 2000.

\end{thebibliography}

\end{document}